\documentclass[nobibnotes,superscriptaddress]{revtex4-1}
\usepackage{epsf}
\usepackage{epsfig}
\usepackage{amsmath}
\usepackage{amsfonts}
\usepackage{amssymb}
\usepackage{graphicx}
\usepackage{multirow}
\usepackage{graphicx}
\usepackage[table,xcdraw]{xcolor}
\usepackage{epsfig}
\usepackage{color}
\usepackage{bm}
\usepackage[yyyymmdd]{datetime}
\usepackage{dashrule}
\usepackage{ulem}

\begin{document}

\title{Hopping induced ground-state magnetism in 6H perovskite iridates}

\author{A. Nag}
\affiliation{Department of Materials Science, Indian Association for the Cultivation of Science, Jadavpur, Kolkata~700032, India}
\affiliation{Current address: Diamond Light Source, Harwell Campus, Didcot OX11 0DE, UK}
\author{S. Bhowal}
\affiliation{School of Physical Sciences, Indian Association for the Cultivation of Science, Jadavpur, Kolkata~700032, India}
\affiliation{Current address: Department of Physics \& Astronomy, University of Missouri, Columbia, Missouri 65211, USA}
\author{M. Moretti Sala}
\affiliation{ESRF-The European Synchrotron, 71 Avenue des Martyrs, 38000 Grenoble, France}
\affiliation{Dipartimento di Fisica, Politecnico di Milano, P.zza Leonardo da Vinci 32, I-20133 Milano, Italy}
\author{A. Efimenko}
\affiliation{ESRF-The European Synchrotron, 71 Avenue des Martyrs, 38000 Grenoble, France}
\author{I. Dasgupta}
\affiliation{School of Physical Sciences, Indian Association for the Cultivation of Science, Jadavpur, Kolkata~700032, India}
\affiliation{Centre for Advanced Materials, Indian Association for the Cultivation of Science, Jadavpur, Kolkata~700032, India}
\author{S. Ray}
\email[Email: ]{mssr@iacs.res.in}
\affiliation{Department of Materials Science, Indian Association for the Cultivation of Science, Jadavpur, Kolkata~700032, India}
\affiliation{Centre for Advanced Materials, Indian Association for the Cultivation of Science, Jadavpur, Kolkata~700032, India}


\begin{abstract}
Investigation of elementary excitations has advanced our understanding of many-body physics governing most physical properties of matter. Recently spin-orbit excitons have drawn much attention, whose condensates near phase transitions exhibit Higgs mode oscillations, a long-sought physical phenomenon [Nat. Phys. {\bf 13}, 633 (2017)]. These critical transition points resulting from competing spin-orbit coupling (SOC), local crystalline symmetry and exchange interactions, are not obvious in Iridium based materials, where SOC prevails in general. Here, we present results of resonant inelastic x-ray scattering on a spin-orbital liquid Ba$_3$ZnIr$_2$O$_9$ and three other 6H-hexagonal perovskite iridates which show magnetism, contrary to non-magnetic singlet ground state expected due to strong SOC. Our results show that substantial hopping between closely placed Ir$^{5+}$ ions within Ir$_2$O$_9$ dimers in these 6H-iridates, modifies spin-orbit coupled states and reduces spin-orbit excitation energies. Here, we are forced to use at least a two-site model, to match the excitation spectrum going in line with the strong intra-dimer hopping. Apart from SOC, low energy physics of iridates is thus critically dependent on hopping, and may not be ignored even for systems having moderate hopping, where the excitation spectra can be explained using an atomic model. SOC which is generally found to be 0.4-0.5~eV in iridates, is scaled in effect down to $\sim$0.26~eV for the 6H-systems, sustaining the hope to achieve quantum criticality by tuning Ir-Ir separation.
\end{abstract}

\maketitle
\noindent
Competition between lattice, spin and orbital degrees of freedom is often responsible for remarkable properties in condensed-matter systems~\cite{rau2016arc}. While atomic spin-orbit coupling is quenched in light transition metal oxides, it starts influencing physical properties in 4$d$ or 5$d$ systems~\cite{kusch2018prb,nag2018prbbyio,agrestini2018prb}. In these materials, spin-orbit coupling (SOC or $\lambda$) manifests itself by splitting wide $d$ bands and effectively enhancing electronic correlations resulting in Mott-insulating states akin to 3$d$ oxides~\cite{kim2008prl,imada1998rmp}. A magnetic condensation of excitations across spin-orbit-coupled states (SOC-states) in some of these Mott-insulators ($d^4$ systems, whose ground states are expected be non-magnetic SOC $J$=0 ($J_0$) singlets, with $S$,$L$ nominally equal to one),  was predicted by Khaliullin~\cite{khaliullin2013prl}, of which Ca$_2$RuO$_4$ has been an exemplary case~\cite{jain2017np}. With the free-ion SOC itself being small in the 4$d^4$ oxides like Ca$_2$RuO$_4$~\cite{jain2017np, souliou2017prl}, SOC-states are readily perturbed by non-cubic crystal-fields ($\Delta_{\textup{CF}}^{\textup{NC}}$) around the 4$d^4$ ions, allowing access to the magnetic $J$=1 ($J_1$)  triplet states. In Ir$^{5+}$ systems however, the prediction seems to have stayed experimentally unclear and debated on account of stronger SOC that raises the excitation energies (0.37~eV) compared to the inter-site exchange interactions~\cite{kusch2018prb,nag2018prbbyio}. In addition, it has been shown recently that even a large $\Delta_{\textup{CF}}^{\textup{NC}}$ (0.325~eV), fails to compete with strong SOC to close the spin-orbit excitation gap comprehensively~\cite{agrestini2018prb}.

\begin{figure*}[tbp]
	\begin{center}
		\includegraphics[width=175mm]{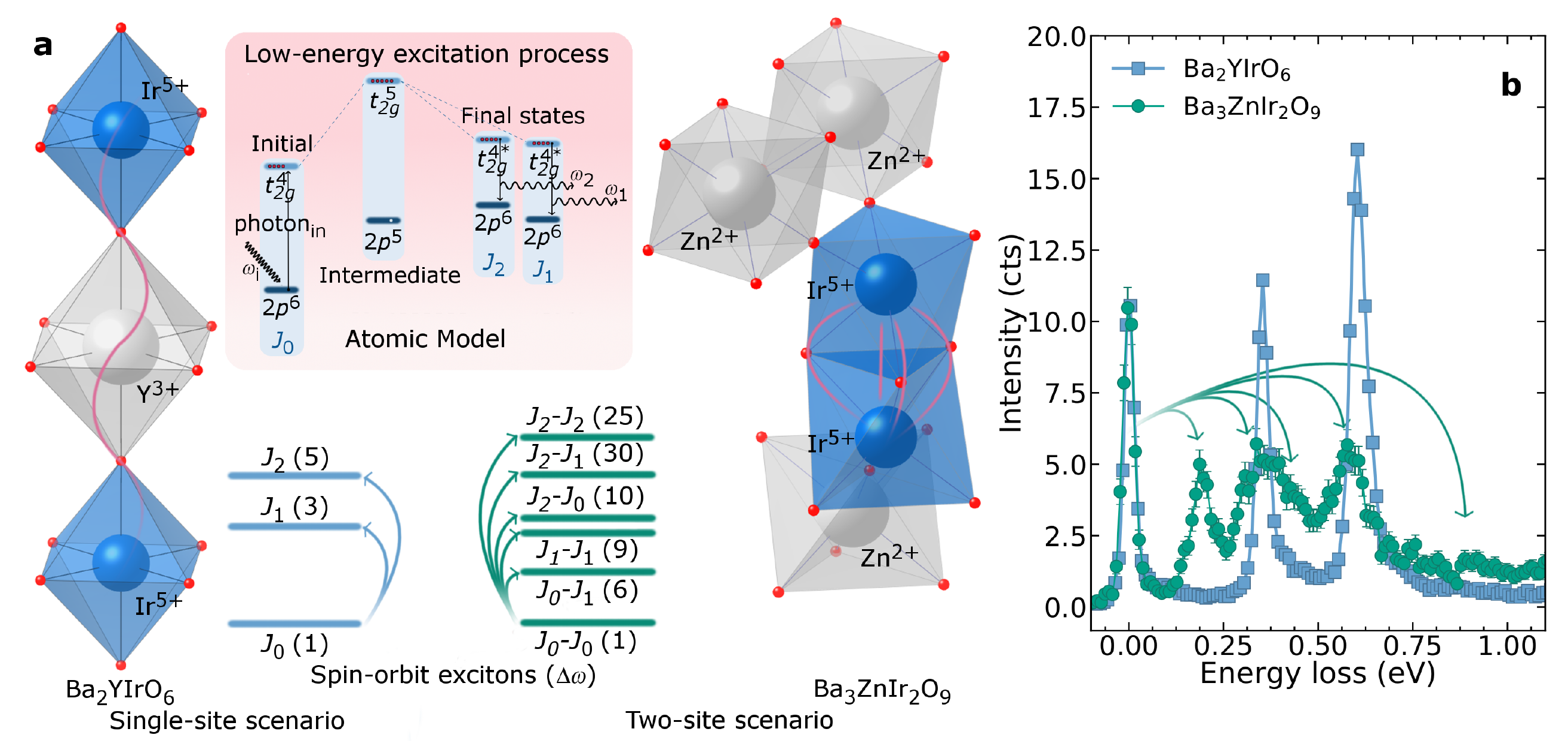}
		\caption{
			Enhanced inter-site hopping rendering possibility of magnetic condensation in 6H-iridates. Inset: Spin-Orbit excitonic process occurring in RIXS (represented by curved arrows in main panel) shown schematically. SOC-states higher in energy are not shown for clarity as transitions to them have low spectral intensities. See text for details. (a) As one goes from double perovskites like Ba$_2$YIrO$_6$, where the Ir$^{5+}$ ions are far apart to 6H-hexagonal perovskites like Ba$_3$ZnIr$_2$O$_9$, where the Ir$^{5+}$ ions form a dimer~\cite{momma2011jac}, inclusion of effective inter-site hopping creates two-site SOC-states ($J_0$-$J_0$, $J_0$-$J_1$, etc.). Panel (b) compares the low-energy RIXS spectra from the SOC-states in Ba$_2$YIrO$_6$ to Ba$_3$ZnIr$_2$O$_9$ showing energy-loss peaks corresponding to excitations to the SOC-states. Presence of additional inelastic peaks and spectral-weight transfer to lower energies make it impossible to describe Ba$_3$ZnIr$_2$O$_9$ using the atomic model.
		} \label{fig1}
	\end{center}
\end{figure*}

Interestingly, in these systems either the Ir$^{5+}$ ions are far apart intervened by closed shell ions (Fig.~\ref{fig1})~\cite{kusch2018prb,nag2018prbbyio}, or are within corner/edge-shared octahedral geometries~\cite{du2013epl}. In contrast, significant magnetic responses are obtained from 6H-hexagonal perovskite iridates Ba$_3$$M$Ir$_2$O$_9$, ($M$~=~Mg, Zn, Ca, and Sr), where the Ir$^{5+}$ ions form face-sharing octahedral dimeric units (Fig.~\ref{fig1}(a))~\cite{nag2016prl,sakamoto2006jssc,nag2018prbbmio}. Since each Ir$^{5+}$ ion in such a situation `sees' another Ir$^{5+}$ ion as its nearest neighbour, hopping between them via O$^{2-}$ ions becomes important~\cite{nag2018prbbmio,kugel2015prb}. In addition to these three Ir-O-Ir superexchange pathways, the small intra-dimer Ir-Ir distances (about 2.7\AA)~\cite{nag2018prbbmio} provide for direct exchanges between the Ir$^{5+}$s. For example, bandwidths ($W=2zt_\textup{eff}$) obtained from density functional theory (DFT) without the influence of SOC, give the effective hopping $t_\textup{eff}$ to be 0.06~eV in double perovskite (DP) Ba$_2$YIrO$_6$ while in 6H Ba$_3$ZnIr$_2$O$_9$ it is around 0.31~eV (see Supplementary Material (SM))~\cite{SM}.

Resonant Inelastic X-Ray Scattering (RIXS), a second order ``photon-in, photon-out'' spectroscopic technique has emerged to be particularly suited to study the hierarchy of these SOC-states in iridates~\cite{ament2011prbr,sala2014prl,kim2017prb,paremkanti2018prb}. The process is sensitive to the changes in energy, momentum and polarization of inelastically scattered x-ray photons representing elementary excitations in a material. With the development of high-resolution spectrometers~\cite{sala2018jsr}, it is now possible to resolve low-energy excitations to such SOC-states in a RIXS experiment~\cite{kusch2018prb,yuan2017prb}. The low-energy spin-orbit excitonic process for an atomic scenario is depicted in the inset of Fig.~\ref{fig1}(a) and explained below. Strictly within an atomic model with strong octahedral crystal-field and SOC, 5$d^4$ electrons in $t_{2g}$ orbitals have their spins $S$=1 coupled to orbital angular momenta $L$=1, forming nominally non-magnetic SOC-states of $J$=0 ($J_0$). The incoming photons with energy $\omega_i$ excite the system from  $\vert i\rangle = \vert 2p^6 t_{2g}^4\rangle$ initial state to an intermediate $\vert n\rangle = \vert 2p^5 t_{2g}^5\rangle$ state with core-holes. This excited state then decays into a final state $\vert f\rangle=\vert 2p^6 t_{2g}^{4\ast}\rangle$ different from the initial state in terms of the arrangement of the SOC electrons ($\frac{6!}{4!2!}-1$ possibilities), and in the process emits photons of energy $\omega_1$,$\omega_2$ and so on. The energy losses $\Delta\omega=\omega_{\rm i}-\omega_1,\omega_2$, then correspond to the energy costs for the different arrangements of the electrons within the SOC $t_{2g}$ orbitals.
Even though RIXS has already been employed to study spin-orbit excitations in DP Ir$^{5+}$ systems, marginal inter-site hopping has had an undetectable effect on the RIXS spectra which could be interpreted by a bare atomic model~\cite{nag2018prbbyio,kusch2018prb,yuan2017prb,SM}. In contrast, we here show using RIXS that enhancement in inter-site hopping due to close placement of Ir$^{5+}$ ions drives the 6H-iridates (Ba$_3$$M$Ir$_2$O$_9$  ($M$~=~Mg, Zn, Ca, and Sr)) into a regime where the effective SOC strength is weakened and the ground state deviates from non-magnetic singlets  (Fig.~\ref{fig1}). This is the first observation of spin-orbit excitons having such low energies in Ir$^{5+}$ systems~\cite{kusch2018prb,nag2018prbbyio}. Since an atomic model fails~\cite{SM}, particularly to explain a low-energy peak around 0.2~eV and  a two-peak feature around 0.4~eV as shown in Fig.~\ref{fig1} (b), we implement exact diagonalisation of a minimal two-site model including the influences of physical parameters: $\lambda$, $\Delta_{\textup{CF}}^{\textup{NC}}$, Hund’s coupling $J_\textup{H}$, and intra-dimer hopping ($t_\textup{dim}$), to map the low-energy excitations in RIXS spectra. We find that the observed features are characteristic of a two-site model with a finite inter-site hopping (see Fig. \ref{fig1}). This is not surprising given the fact that the Ir-ions in these 6H-iridates interact strongly with their nearest neighbours forming Ir-Ir dimers~\cite{nag2016prl,nag2018prbbmio}, in comparison to DP iridates where the Ir-ions have little interaction. Strangely however, we find that $\lambda$, usually considered to be 0.4-0.5~eV for iridates, is scaled down effectively to $\sim$0.26~eV in the presence of strong $t_\textup{dim}$. Moderate hopping in other systems like Ba$_2$YIrO$_6$, may similarly be suspected to rescale the atomic $\lambda$, even though imperceptible in RIXS~\cite{nag2018prbbyio}.

\begin{figure*}[tbp]
	\begin{center}
		\includegraphics[width=175mm]{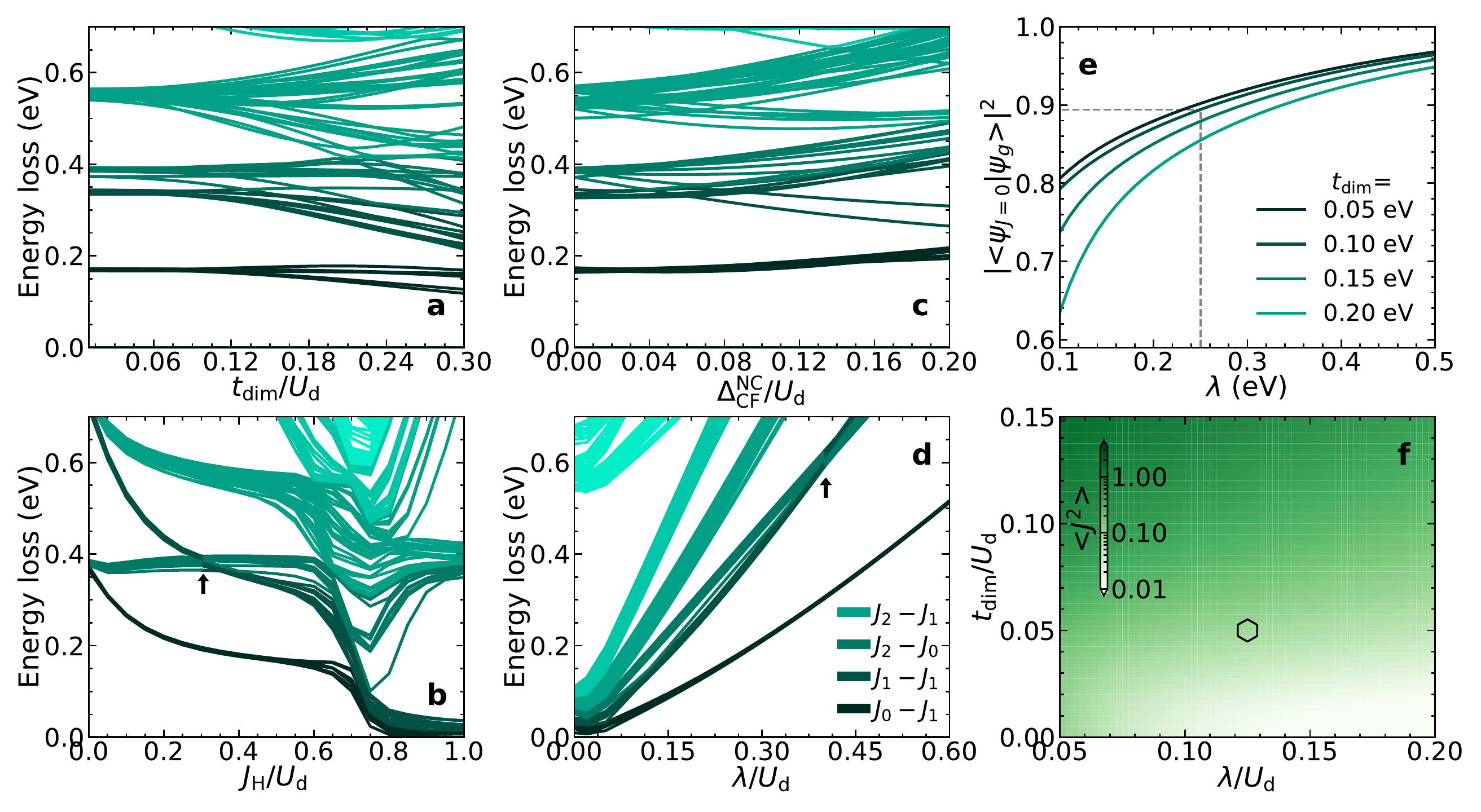}
		\caption{
			Panels (a)-(d) show the effects of varying physical parameters $t_\mathrm{dim}/U_d$,  $J_\mathrm{H}/U_d$, $\Delta _{\textup{CF}}^{\textup{NC}}/U_d$ and $\lambda/U_d$ respectively, on the SOC excitonic gaps, evaluated using the two-site model Hamiltonian given in equation~(\ref{multiplet_method}). Each set of SOC-states is named after their parent atomic SOC-states, and coloured with a different shade as labelled in panel (d). Panel (e) shows that for $\lambda$~=~0.25 eV (relevant for the 6H-iridates), the two-site ground state ($\psi_g$) deviates from $J~=~0$ state by about 10\%. The phase diagram in panel (f) shows the variation of the $<J^2>$ expectation value in the parameter space of $t_\mathrm{dim}/U_d$ and $\lambda/U_dS$. The mark denotes the estimated $J\neq0$ scenario in the 6H-iridates. Parameters of the Hamiltonian are (unless it's variation has been shown): $t_{\rm dim}$ = 0.1 eV, $\Delta_\textup{CF}^\textup{NC}$ = 0.02 eV, $J_\textup{H}$ = 0.5 eV, $\lambda$ = 0.25 eV and $U_d$ = 2.0 eV.
		} \label{fig2}
	\end{center}
\end{figure*}

In order to characterise the spin-orbit excitons in presence of intra-dimer hopping (beyond capacity of single-particle theories), we have employed many-body multiplet formalism incorporated within the following two-site model Hamiltonian for the Ir-$t_{2g}$ orbitals~\cite{matsuura2013jpsj,bhowal2018prb}:

\begin{equation}\label{multiplet_method}
H = \sum_{\substack{i=1,2}} (H_i^{\Delta_\mathrm{CF}^\mathrm{NC}}+ H_i^\mathrm{int}+H_i^\mathrm{SO})+ H^{t},
\end{equation}

where $H_i^{\Delta_\textup{CF}^\textup{NC}}$, $H_i^\textup{int}$ and $H_i^\textup{SO}$ are the respective Hamiltonians representing the non-cubic crystal-field, Coulomb interactions and SOC for the $i^{\rm th}$ site. $H_i^\textup{int}$, known as Kanamori Hamiltonian, includes intra- ($U_d$) and inter-orbital ($U_d^\prime$) Coulomb interactions alongside Hund's coupling ($J_\textup{H}$), and are related by $U_d=U_d'+2J_\textup{H}$ (see SM for detailed expressions)~\cite{SM}. In addition to these on-site terms, we introduce $H^{t}$ to account for inter-site hopping $t_{ij}^{l\sigma,m\sigma^\prime}$ between the three $t_{2g}$ orbitals at the two Ir$^{5+}$ sites $i$ and $j$.
While within the atomic model~\cite{SM}, SOC gives rise to multiplet states $J_0$, $J_1$, $J_2$ ... (see Fig.~\ref{fig1} (a)), in presence of strong $t_{\rm dim}$ these SOC-states are influenced by neighbouring atomic multiplet states forming two-site SOC-states (shown schematically in Fig. \ref{fig1} (a)). The spin-orbit exciton energies can be obtained from the difference in energy eigenvalues calculated using exact diagonalization of the two-site model. Due to the inclusion of inter-site hopping in the two-site model, new sets of SOC-states appear out of the interaction between the atomic SOC-states at the two sites: $J_0$-$J_0$ (1), $J_0$-$J_1$ (6), $J_1$-$J_1$ (9), $J_0$-$J_2$ (10),$J_1$-$J_2$ (30),$J_2$-$J_2$ (25) etc. The new sets of two-site SOC-states are denoted according to their parent atomic SOC-states ($J_0$-$J_0$, $J_0$-$J_1$, etc.) and the numbers within the parentheses indicate the number of SOC-states having the same origin.  Excitations from the $J_0$-$J_0$ SOC-states to the sets originating from the interaction between atomic SOC-states $J_1$-$J_1$ or $J_0$-$J_2$, form the inelastic features in the RIXS spectra. The fact that strong intra-dimer hopping plays a key role in modifying low-energy SOC excitation spectrum, is directly verified from the experimental results (Fig. \ref{fig1} (b)).

\begin{figure*}[tbp]
	\begin{center}
		\includegraphics[width=175mm]{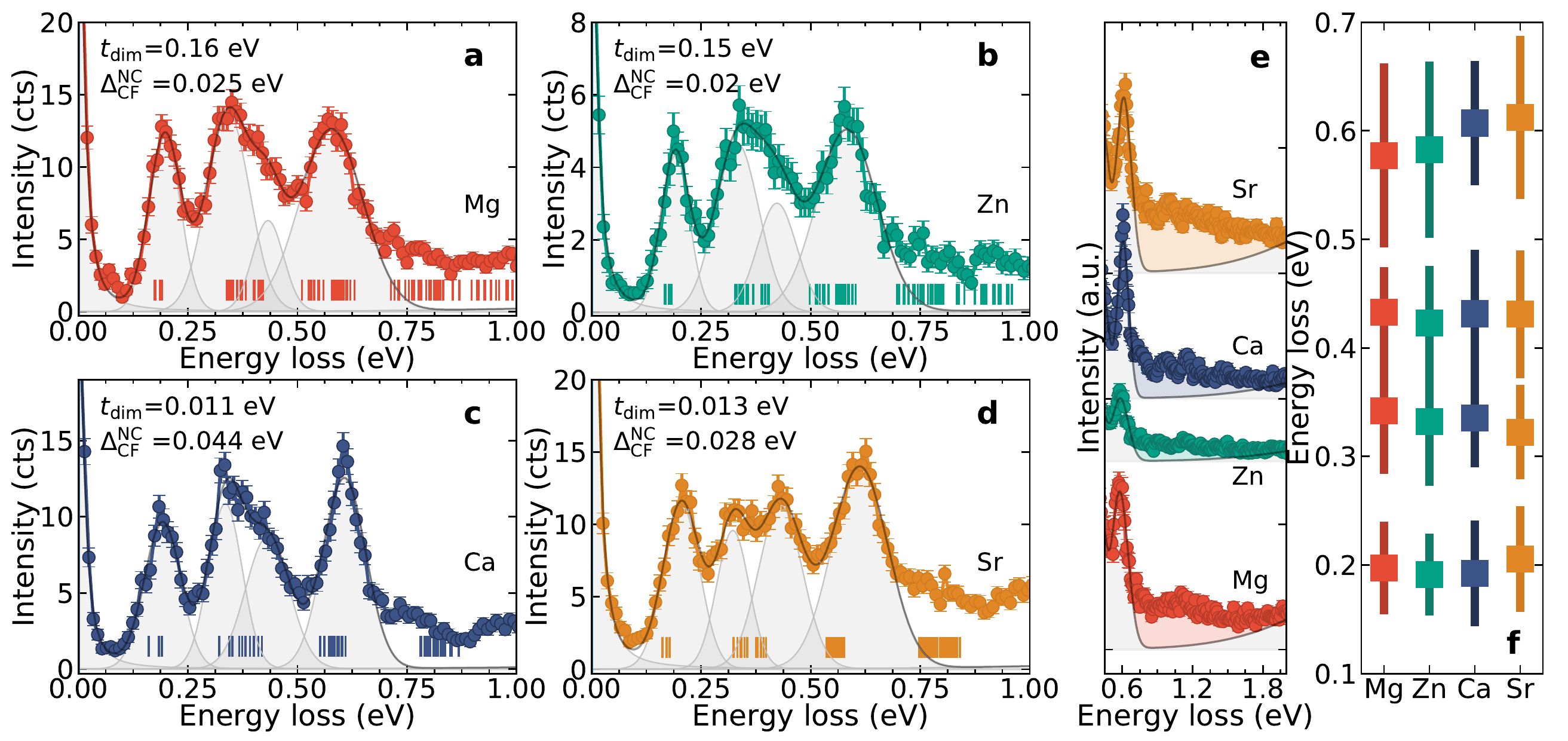}
		\caption{Low-energy RIXS spectra from the 6H-iridates and energy-losses corresponding to the two-site SOC-states. Each spectrum is decomposed into four sets of spin-orbit excitations by fitting Gaussian line-shapes (shaded grey) in panels (a-d). Vertical bars represent energies of two-site SOC-states evaluated using $J_\textup{H}$ = 0.5 eV and $\lambda$ = 0.25 eV in equation~(\ref{multiplet_method}) and parameters given in each panel. In panel (e) the spectra are shifted vertically to show continua of low intensity excitations above the well defined peaks (shaded regions between the spectra and Gaussian tails from $e_g$ peaks). Fitted peak-positions (squares) and peak-FWHMs (length of the vertical lines through the squares) for the four 6H-iridates shown in panels (a-d), are presented in panel (f).
		} \label{fig3}
	\end{center}
\end{figure*}
In our calculation, we have assumed the hopping to be diagonal in the spin-space {\it i.e.,} $t_{\rm dim} = t_{ij}^{l\sigma,m\sigma}$. Also we have supposed that the hopping parameters are diagonal and equal in the orbital space to illustrate the effect of different parameters on the energies of the spin-orbit excitons (Fig.~\ref{fig2}). A trigonal crystal-field is also assumed in this case. However, more realistic hopping and non-cubic matrices guided by the first principles calculations are considered to fit the RIXS spectra of the 6H-iridates. It is important to note that the non-cubic crystal-field and the hopping terms take care of the non-local effects within the solid. Such a simple $t_{2g}$-only two-site model captures the essential features of these 6H-iridates owing to the dimeric interaction and large $t_{2g}$-$e_g$ splitting (allowing us to neglect the $e_g$ orbitals (see SM Fig.~2))~\cite{SM}. Taking into account the $\Delta_\textup{CF}^\textup{NC}$ and $t_\textup{dim}$ (= $t_{ij}^{l\sigma,m\sigma^\prime}$), which are both set-up in part by O$^{2-}$ ions, also indirectly includes their effects in the model. Fig.~\ref{fig2} shows the influence of each of these physical parameters on the excitonic gaps within the two-site model calculation. In contrast to the atomic states, spin-orbit excitation energies obtained from the two-site model also depend on Coulomb repulsion $U_d$ since the superexchange energy varies as $\sim t_{\rm dim}^2/U_d$ in the latter case. It is clear from Fig.~\ref{fig2}(a-d) that $\Delta_\textup{CF}^\textup{NC}$ and $t_\textup{dim}$ control spread of each set of states, while their mean energies are primarily dictated by  $J_\textup{H}$ and $\lambda$. An interesting feature in the spin-orbit exciton energy spectrum is crossover between $J_0$-$J_2$ and $J_1$-$J_1$ states as we go from $J_\textup{H} << \lambda$ to the $J_\textup{H} >> \lambda$ limit~\cite{SM}. Even though this two-site model is a simplistic representation of processes in a real solid, it captures the salient features of the low-energy RIXS spectra of the 6H-perovskite iridates as shown ahead. Most importantly, as seen from Fig.~\ref{fig2}(e) that for $\lambda$~=~0.25 eV (relevant for the 6H-iridates), the two-site ground state deviates from $J~=~0$ state by about 10\% and as a consequence expectation value  $<J^2>$ of the ground state attains a non-zero value and hence small moments.

\begin{figure*}[tbp]
	\begin{center}
		\includegraphics[width=175mm]{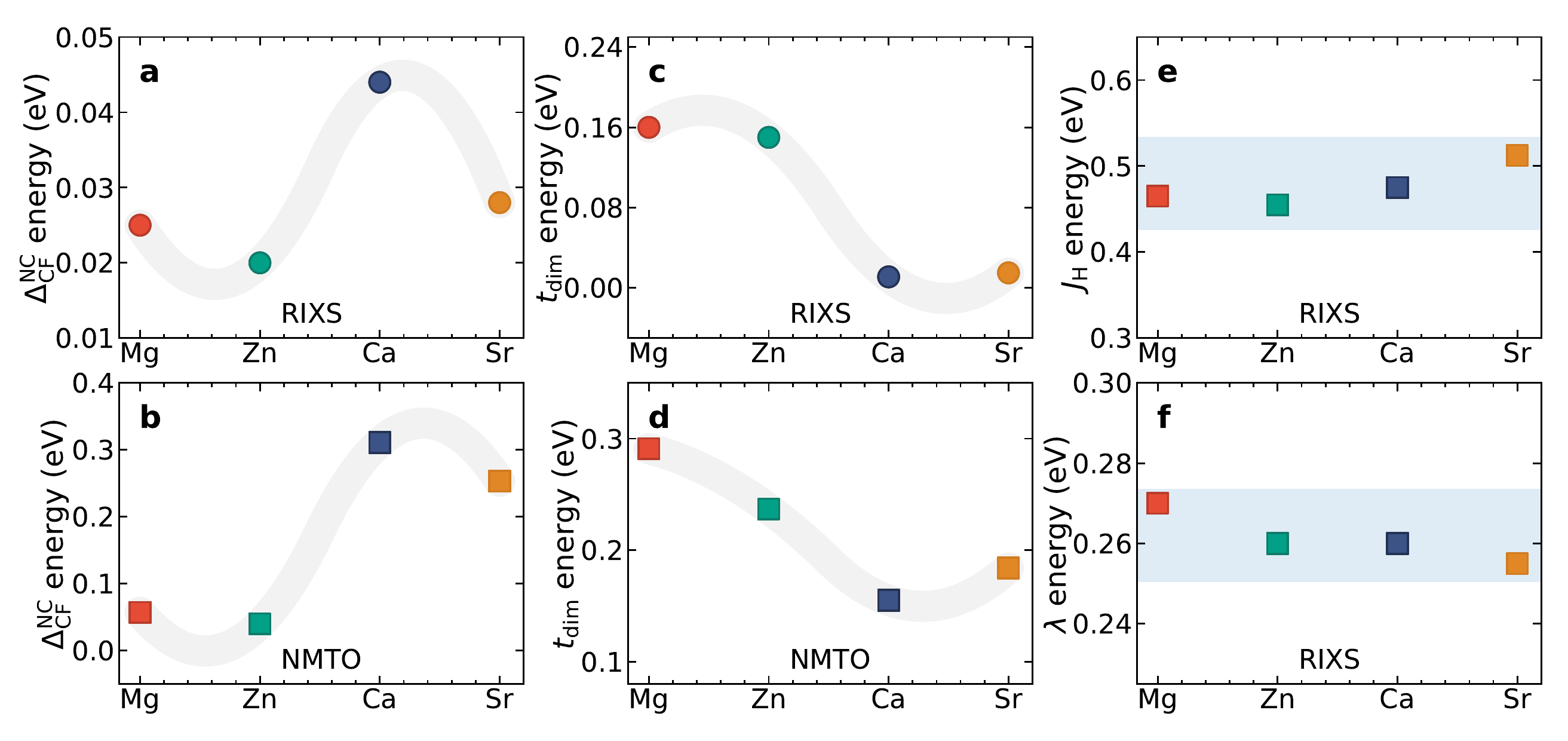}
		\caption{
			Comparison of physical parameters obtained from DFT to the ones extracted by fitting RIXS spectra with two-site model Hamiltonian for the 6H-iridates. Energy-loss values associated with SOC-states were calculated using the two-site model Hamiltonian (equation~(\ref{multiplet_method})) and matched to experimentally obtained inelastic features, by varying $\Delta _{\textup{CF}}^{\textup{NC}}$, $t_\mathrm{dim}$, $J_\mathrm{H}$ and $\lambda$. Maximum components of $\Delta _{\textup{CF}}^{\textup{NC}}$, $t_\mathrm{dim}$ matrices for each system giving best fits to RIXS are shown in panels (a) and (c) respectively (see text). Plotted in panels (b) and (d) are the corresponding values obtained from DFT calculations using NMTO. The points are connected by shaded splines of degree 2 to emphasise similar variation in values obtained using the two methods. Panels (e) and (f) show average values of $J_\mathrm{H}$ and $\lambda$ obtained for the best fits to RIXS spectra in each system.
		} \label{fig4}
	\end{center}
\end{figure*}

High-resolution (0.029~eV at Ir-$L_3$-edge) RIXS spectra of the 6H-iridates (Ba$_3$$M$Ir$_2$O$_9$, $M$~=~Mg, Zn, Ca, and Sr) collected at 20~K at the ID20 beamline of ESRF – The European Synchrotron at 20~K~\cite{sala2018jsr}, are shown in Fig.~\ref{fig3}(a-d), all of which show similar low-energy inelastic features unlike the RIXS from DPs like Ba$_2$YIrO$_6$ (See Fig.~\ref{fig1}(b)) exhibiting the strong influence of intra-dimer hopping in these systems. The inelastic features of each spectrum, are fitted with four Gaussian line-shapes representing excitations from the $J_0-J_0$ state to higher lying two-site SOC states (see SM for details)~\cite{SM}. Since excitations to the set of SOC-states $J_1-J_1$ and $J_2-J_0$ are not clearly resolved (strongest splitting is seen for Ba$_3$SrIr$_2$O$_9$, Fig.~\ref{fig3}(d)), the fits to these two sets of peaks are done by limiting their Gaussian variances to 0.05~eV. A Gaussian tail from the high-energy $t_{2g}$-$e_{g}$ excitation peak is also included in the fitting, although its effect on the 4-SOC peaks is minimal (Fig.~\ref{fig3}(e)). Although, excitations within the higher-energy states cannot be ruled out given their strong intermixing, these excitations would have comparatively weak intensities and are not resolved in the present experiments. A broad continuum of excitations can be seen (Fig.~\ref{fig3}(e)) above 0.7~eV in all the systems which may be related to low-intensity transitions to $J_2-J_2$ states besides multiparticle excitations~\cite{kim2017prb,paremkanti2018prb}. We note that the spectral weights of the $J_2-J_2$ excitation features ($\sim 0.75$~eV) are substantially suppressed in spite of large number of states for all the 4 systems again showing significant hopping induced deviation of these states from the pure atomic limit (see SM Fig.~5))~\cite{SM}. 

The peak positions and their widths extracted from the fitting of the energy-loss features are shown in Fig.~\ref{fig3}(f). The slight variation in the peak positions and widths are related to the details of the local geometry of these four iridates. This becomes evident when we try to place the difference in eigenvalues obtained from the exact diagonalization of the two-site model Hamiltonian given in equation (\ref{multiplet_method}) with different parameters, within the FWHMs of these peaks (SM Fig.~4)~\cite{SM}.

For the two smaller cations Mg$^{2+}$ (0.72 \AA) and Zn$^{2+}$, (0.74 \AA) the compounds Ba$_3$MgIr$_2$O$_9$ and Ba$_3$ZnIr$_2$O$_9$ stabilize in a $P{\rm 6}$$_{3}$/$mmc$ space group, while Ca$^{2+}$ (1.00 \AA) and Sr$^{2+}$ (1.18 \AA) allow formation of Ba$_3$CaIr$_2$O$_9$ and Ba$_3$SrIr$_2$O$_9$ only within a lower symmetry space group $C2/c$~\cite{nag2016prl,nag2018prbbmio,sakamoto2006jssc,shannon1976ac,SM}. Each of these 6H-iridates therefore possesses slightly modified IrO$_6$ octahedral environment resulting in different  $t_{\rm dim}$ and $\Delta_{\textup{CF}}^{\textup{NC}}$ values. We find that RIXS spectra for the 6H-iridates can be fitted over a small range of values for all the parameters (SM Table II)~\cite{SM}. A realistic estimation of hopping and non-cubic crystal-field energies for the different local geometries of the Ir$^{5+}$ ions in the 6H-iridates is obtained by extracting the low-energy tight-binding model retaining only Ir-$t_{2g}$ in the basis from the muffin-tin orbital (MTO) based N$^{th}$ order MTO (NMTO) method ~\cite{NMTO1,NMTO2,NMTO3} as implemented in the Stuttgart code. While $P{\rm 6}$$_{3}$/$mmc$ symmetry of Ba$_3$MgIr$_2$O$_9$ and Ba$_3$ZnIr$_2$O$_9$ leads to trigonal distortion of the IrO$_6$ octahedra so that Ir-$t_{2g}$ orbitals split into singly degenerate $a_{1g}$ and $e_{g}^\pi$ states,  monoclinic distortions in Ba$_3$CaIr$_2$O$_9$ and Ba$_3$SrIr$_2$O$_9$ remove completely the degeneracy of the $t_{2g}$ states and is reflected accordingly in the non-cubic crystal field matrices (see SM Section V). This results in diagonal hopping matrices for Ba$_3$MgIr$_2$O$_9$ and Ba$_3$ZnIr$_2$O$_9$, while Ba$_3$CaIr$_2$O$_9$ and Ba$_3$SrIr$_2$O$_9$ have relatively complicated forms with non-zero off-diagonal elements (see SM Section VI). The values obtained from the first-principles calculations are taken as initial guesses and are then renormalized to fit the RIXS spectra, while keeping the ratios between different elements of the matrices for hopping and non-cubic crystal-field fixed to that obtained from DFT. Parameters of the Hamiltonian which give energy-loss values within the FWHMs of the Gaussian fits to the RIXS peaks are then extracted as shown in SM Fig.~4 (a)-(d). In our calculations, we keep the value of $U_d$ fixed at 2~eV~\cite{kusch2018prb}.

The best matched energy-loss values of the two-site SOC-states calculated using a set of values for the parameters $\Delta_\textup{CF}^\textup{NC}$, $t_\textup{dim}$, $J_\textup{H}$ and $\lambda$ for each of the four systems, are shown as vertical bars in Fig.~\ref{fig3}(a-d). The maximum components of $\Delta_\textup{CF}^\textup{NC}$ and $t_\textup{dim}$ matrices used for the fits are then compared to the corresponding values obtained using NMTO method on these systems in Fig.~\ref{fig4}(a-d). We find that the parameters extracted from RIXS are scaled down w.r.t to the ones obtained from first principles. However, as can be seen from panels of Fig.~\ref{fig4}(a-d), the reduction in values is not arbitrary and is qualitatively similar for the four 6H-iridates. It should be mentioned here, that overestimation of parameters from first principles is known and reported for honeycomb iridates with $d^5$ electronic configuration~\cite{kim2016prl}. The Hund's coupling estimated from RIXS ($\sim 0.45$~eV) is similar for all four (see Fig.~\ref{fig4}(e)), comparable to other 5$d^5$ systems but higher than the 5$d^4$ DP-iridates~\cite{yuan2017prb,kusch2018prb}. However, most striking is the substantially suppressed value of SOC ($\sim 0.26$~eV) compared to any other iridate reported as yet, proving the decisive role of intra-dimer hopping in these systems. In this context, reduced covalency has recently been ascribed to increased SOC in iridium fluorides~\cite{rossi2017prb}. RIXS of these 6H-iridates provide first clear experimental evidences of atomic non-magnetic SOC $J$~=~0 state breakdown, in presence of solid state effects. The two-site model employed may thus be appropriate to describe the low-energy physics of other systems where the Ir-ions are placed close-by in the crystal structure~\cite{takayama2018,streltsov2018prb}. The polycrystalline nature of our samples and two-site theoretical modelling forbid estimation of the momentum dependence of the spin-orbit excitons presently and can be a future direction of work. The identification of inter-site hopping as a critical parameter which can drastically change the effective strength of SOC even so, may be utilised to tune magnetism in iridates.

\section*{Acknowledgements}
SR and IDG thank Technical Research Center of IACS. SR also thanks Department of Science and Technology (DST) [Project No. WTI/2K15/74], UGC-DAE Consortium for Scientific Research, Mumbai, India [Project No. CRS-M-286] for support, and Jawaharlal Nehru Centre for Advanced Scientific Research from DST-Synchrotron-Neutron project, for performing experiments at ESRF (Proposal No. HC-2872).

\setcounter{figure}{0}

\begin{center}
	{\bf \Large {Supplementary information for\\ Hopping induced ground-state magnetism in 6H perovskite iridates} }
\end{center}


\section{Sample synthesis:}Polycrystalline samples of Ba$_3$$M$Ir$_2$O$_9$ ($M$~=~Mg, Zn, Ca, and Sr) were synthesized by standard solid state reaction using stoichiometric amounts of BaCO$_3$, MgO, ZnO, CaCO$_3$, SrCO$_3$, and IrO$_2$ as starting materials~\cite{SI_sakamoto2006jssc,SI_nag2016prl,SI_nag2018prbbmio}. The structural parameters used for the calculations are given in earlier reports~\cite{SI_nag2018prbbmio}.

\section{Resonant Inelastic X-ray Scattering}
Ir-$L_3$-edge RIXS on polycrystalline 6H-iridates were done at the I20 beamline of ESRF – The European Synchrotron at 20~K~\cite{SI_sala2018jsr}. Polarisation of the incident x-ray beam was kept parallel to the horizontal scattering plane and the scattered beam was collected at 90$^\circ$ w.r.t the incident beam. Energy of the incident x-ray was varied across Ir-$L_3$ absorption edge and the corresponding energy-loss spectrum was recorded (see Fig.~1). Low-energy features adjacent to the elastic peaks arising due to excitations across the SOC-states were found to be enhanced at an incident energy of 11.216~keV and was fixed for subsequent high resolution measurements. Low-resolution RIXS (0.280~eV) focussing on high-energy loss features for all the four systems are shown in Fig.~2. The high-energy loss features identified as $t_{2g}$-$e_{g}$ transitions and charge-transfer excitations from O 2$p$ to vacant Ir-orbitals were found to be similar to DP Ba$_2$YIrO$_6$~\cite{SI_sala2014prl,SI_ishii2011prb}. To distinguish the low-energy SOC-states, we collected RIXS spectrum on each sample with a resolution of 0.029~eV.
\begin{figure}[h]
	\begin{center}
		\includegraphics[width=\columnwidth]{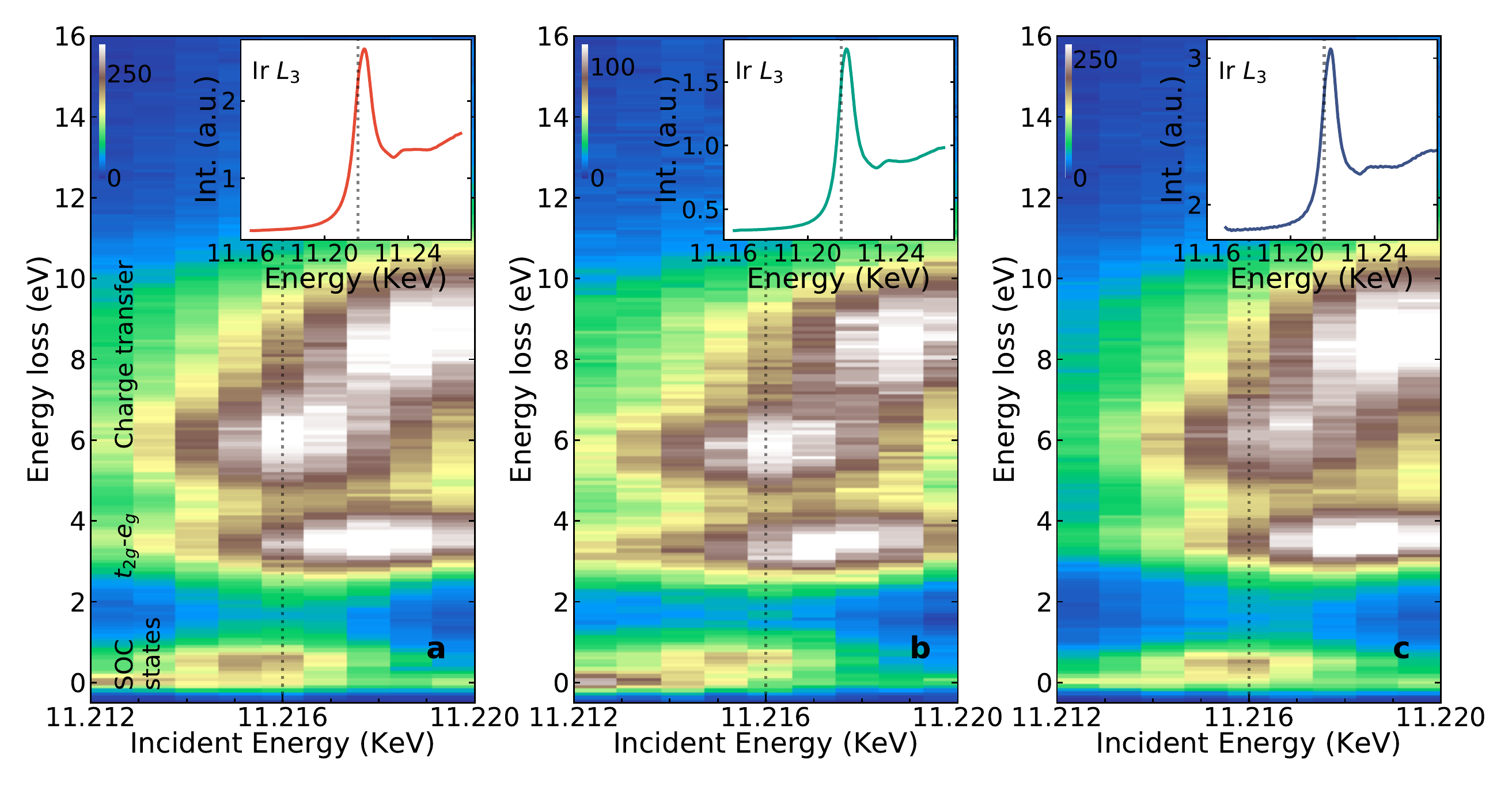}
		\caption{\textbf{Energy dependence of inelastic features in 6H-iridates.}
			Panels (a),(b) and (c) show energy-loss spectra with varying incident energies across Ir-$L_3$ absorption edge (shown as insets) for Ba$_3$MgIr$_2$O$_9$, Ba$_3$ZnIr$_2$O$_9$ and Ba$_3$CaIr$_2$O$_9$ respectively.  Low-energy features adjacent to the elastic peaks arising due to excitations across the spin-orbit coupled $J$ states were found to be enhanced at an incident energy of 11.216~keV and was fixed for subsequent measurements.
		} \label{SIfig1}
	\end{center}
\end{figure}

\begin{figure}[h]
	\begin{center}
		\includegraphics[width=\columnwidth]{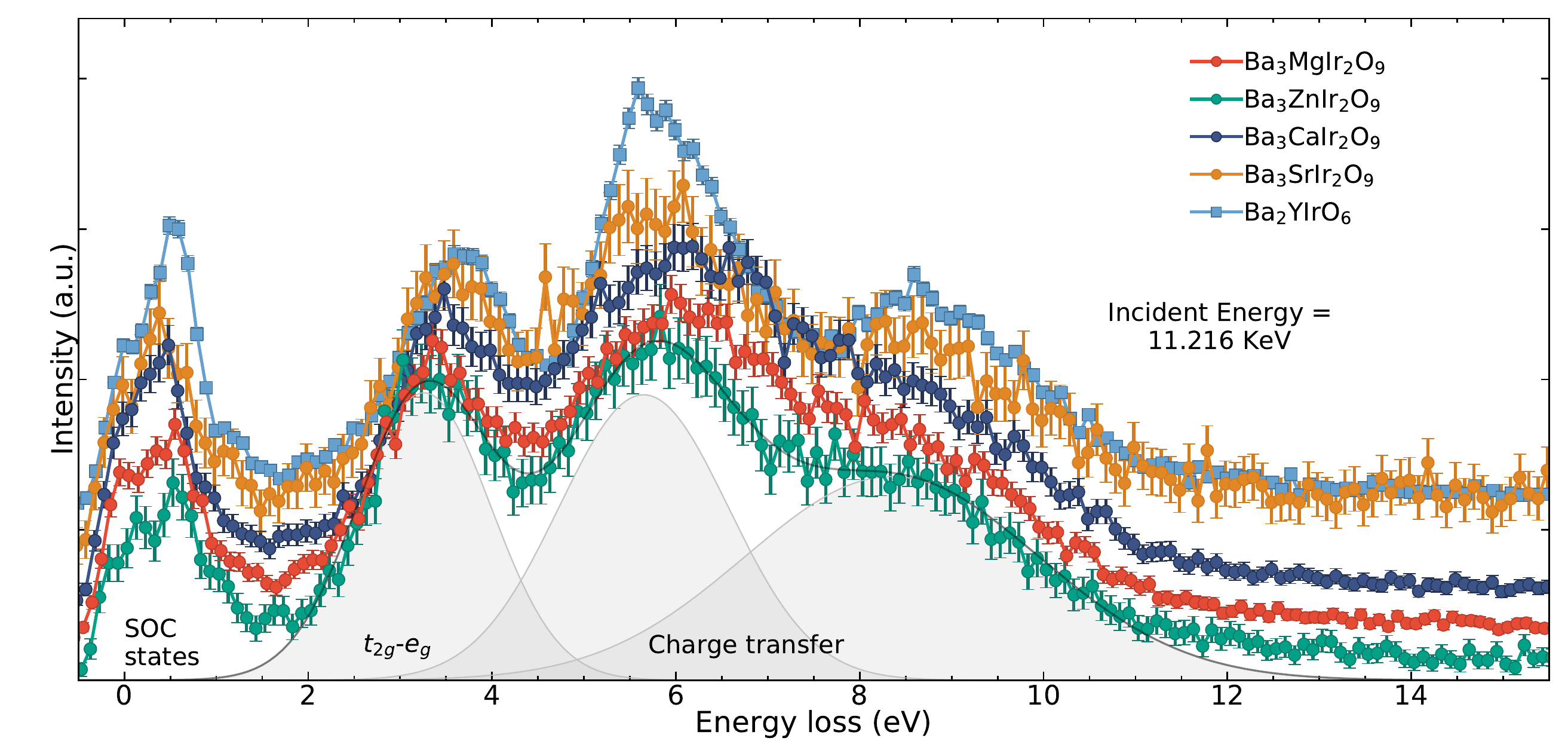}
		\caption{\textbf{Comparison of high-energy-loss features.}
			Vertically shifted low-resolution energy-loss spectra showing the crystal-field and charge-transfer excitations for Ba$_3$$M$Ir$_2$O$_9$ ($M$~=~ Mg, Zn, Ca, Sr), apart from unresolved SOC excitations. For comparison, RIXS spectrum from Ba$_2$YIrO$_6$ is also shown acquired with identical resolution.
		} \label{SIfig2}
	\end{center}
\end{figure}

\section{Atomic model Hamiltonian:}
An atomic many-body Hamiltonian for the $d^4$ configuration \cite{nag2016prl} in  presence of strong SOC gives rise to spin-orbit coupled states: $J_0$ (1), $J_1$ (3), $J_2$ (5), $J_2$ (5), and $J_0$ (1), where numbers within parentheses denote the number of SOC-states having same total angular momentum (see Fig.~1(a) of main text). The energy eigenvalues of these spin-orbit coupled-states are dictated by the interaction parameters such as Hund's coupling $J_\textup{H}$, Coulomb interaction $U_d$ and SOC constant $\lambda$. The energy difference between these states are however independent of Coulomb interaction $U_d$. It is interesting to note that the explicit energies of the  $J_0$ and $J_2$ states depend on the relative strength of the parameters $J_\textup{H}$ and  $\lambda$ keeping the energy difference always fixed at $3\lambda/2$.
In the limit $J_\textup{H} >> \lambda$, $E_{J_0} = E_0 - 4U + 7J_\textup{H} - \lambda$ and $E_{J_2} = E_0 - 4U + 7J_\textup{H} + \lambda/2 $; for $J_\textup{H} << \lambda$, $E_{J_0} = E_0 - 4U + 7J_\textup{H} - 2\lambda$ and $E_{J_2} = E_0 - 4U + 7J_\textup{H} - \lambda/2 $~\cite{SI_kim2016prl}.

Such an atomic many-body Hamiltonian is sufficient to describe the RIXS spectra of the DP iridates \cite{SI_yuan2017prb,SI_nag2018prbbyio,SI_kusch2018prb,SI_paremkanti2018prb} with triplet ($J_0 \rightarrow J_1$) and quintet ($J_0 \rightarrow J_2$) spin-orbit excitons, however it fails to interpret the RIXS spectra for the present 6H-iridates (Ba$_3M$Ir$_2$O$_9$, $M$ = Zn, Mg, Ca and Sr).

\begin{figure}[h]
	\begin{center}
		\includegraphics[width=\columnwidth]{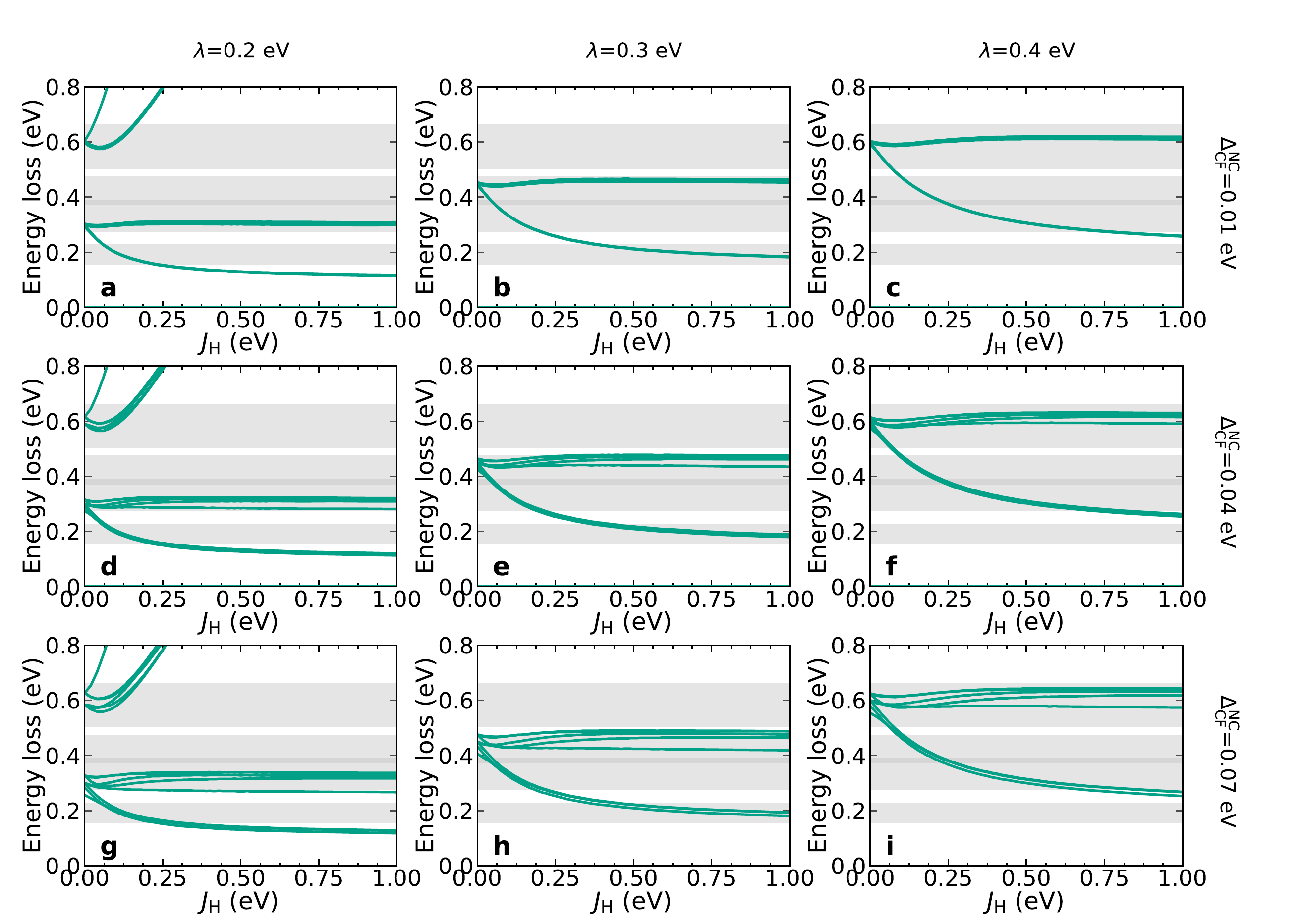}
		\caption{\textbf{Failure of atomic model for 6H-iridates.}
			The energy losses as a function of $J_\textup{H}$ obtained from an atomic model with different $\Delta_\textup{CF}^\textup{NC}$ (panel columns) and $\lambda$ (panel rows) values. Horizontal grey bands are the FWHMs of the Gaussian fits to peaks obtained from RIXS on Ba$_3$ZnIr$_2$O$_9$. It can be seen that for no realistic value of the physical parameters in the atomic model, the energy losses can be matched to that of the experiment. Also, the 4-peak low-energy RIXS spectrum is not replicated for any set of parameters.
		} \label{SIfig3}
	\end{center}
\end{figure}

\section{Two-site model Hamiltonian: }
In order to decipher the low-energy RIXS spectra of these 6H-iridates we consider the following two-site many-body Hamiltonian for the Ir-$t_{2g}$ orbitals
~\cite{SI_matsuura2013jpsj,SI_bhowal2018prb},
\begin{eqnarray}\label{multiplet_method_det}
H &=& \sum_{\substack{i=1,2}} (H_i^{\Delta_\mathrm{CF}^\mathrm{NC}}+ H_i^\mathrm{int}+H_i^\mathrm{SO})+ H^{t}   \\
&=& \sum_{\substack{i=1,2}} \Bigg( \sum_{\substack{l,m,\sigma}}
\epsilon_{lm}d_{l\sigma}^\dagger d_{m\sigma}
+ U_d \sum_{l=1,2,3} n_{l \uparrow} n_{l \downarrow}   \nonumber
+\frac{U_d^\prime-J_d}{2} \sum_{\substack{l,m=1,2,3 \\
		(l \neq m)}}
\sum_{\sigma} n_{l\sigma}n_{m \sigma} \nonumber
+\frac{U_d^{\prime}}{2} \sum_{\sigma \neq \sigma^{\prime}}\sum_{\substack{l,m=1,2,3 \\
		(l \neq m)}}
n_{l\sigma}n_{m\sigma^{\prime}} \nonumber \\
&+& \frac{J_\mathrm{H}}{2} \sum_{\substack{l,m=1,2,3 \\
		(l \neq m)}}
(d_{l \uparrow}^\dagger d_{m \uparrow}d_{l \downarrow}^\dagger d_{m \downarrow}+ h.c.)
+   \frac{i\lambda}{2} \sum_{\substack{lmn \\
		\sigma \sigma^\prime}} \epsilon_{lmn} \sigma_{\sigma \sigma'} ^n d_{l\sigma}^\dagger d_{m\sigma'} \Bigg)
+ \sum_{\substack{i \ne j}} \sum_{\substack{l,m=1,2,3}} \sum_{\substack {\sigma, \sigma^\prime}} t_{ij}^{l\sigma,m\sigma^\prime}d_{il\sigma}^\dagger d_{jm\sigma^\prime},
\end{eqnarray}

where $i$,$j$ are site indices, $l$, $m$ are orbital indices and $\sigma, \sigma'$ are spin indices. $H_i^{\Delta_\textup{CF}^\textup{NC}}$, $H_i^\textup{int}$ and $H_i^\textup{SO}$ are the respective Hamiltonians representing the non-cubic crystal-field, Coulomb interactions and SOC for the $i^{\rm th}$ site. The last term represents the hopping between two different sites $i$ and $j$. The parameters $\epsilon_{lm}$, $U_d$, $U_d^{\prime}$, $J_\textup{H}$, $\lambda$ and
$t_{ij}^{l\sigma,m\sigma^\prime}$ are respectively the on-site energies of the $t_{2g}$ orbitals, intra and inter-orbital Coulomb interaction, Hund's coupling, SOC and the hopping between nearest neighbour sites $i$ and $j$.

\begin{figure}[]
	\begin{center}
		\includegraphics[width=\columnwidth]{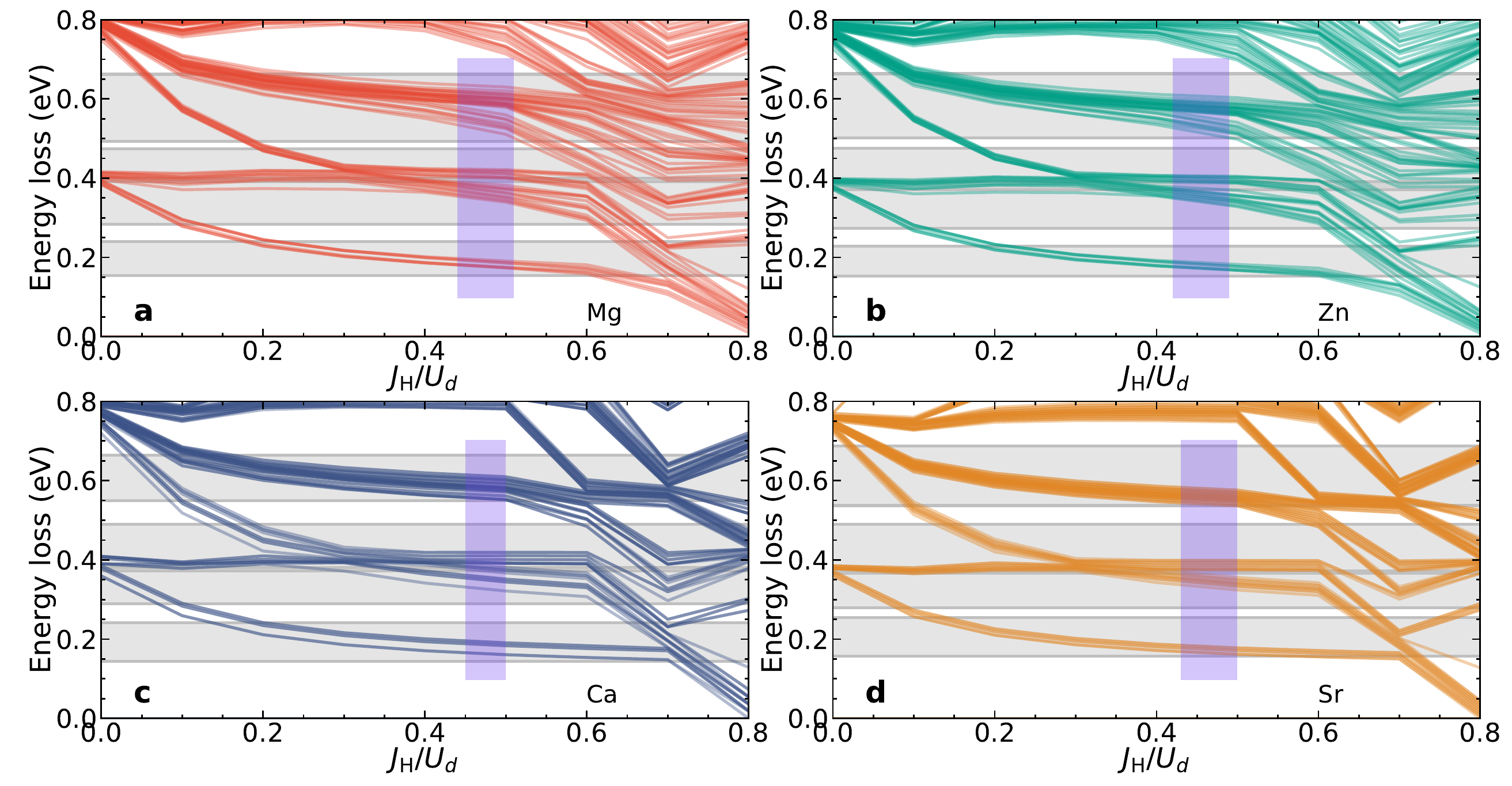}
		\caption{\textbf{Fitting of the RIXS results with the two-site model.}
			Horizontal grey bands are the FWHMs of the peaks obtained from RIXS on Ba$_3$$M$Ir$_2$O$_9$ ($M$=Mg,Zn,Ca,Sr). Calculated energy differences between the SOC-states, according to equation (1) of main text are shown as continuous lines for the 4 systems. Vertical band in each panel represents the Hund's coupling $J_{\textup{H}}$, for which the calculated energy-loss values of each set of two-site SOC-states, lie within the FWHMs of experimentally observed peaks. $J_\textup{H}$ values even lower than the shaded ones may seem to satisfy the experimental results, however it is only within the chosen shaded region that we can observe a separation of states within the energy-loss region 0.3-0.5 eV, resembling the 4-peak low-energy RIXS spectra. Other parameter values used to calculate the theoretical energy-losses in the two-site model are given in italics in Table~\ref{rixsfit}.
		}\label{SIfig4}
	\end{center}
\end{figure}

A crossover in energy between $J_1$-$J_1$ and $J_0$-$J_2$ SOC-states can be seen depending on the relative strength of the $J_\textup{H}$ and $\lambda$  (marked by arrows in Fig. 2(b,d) of the main paper). This is due to the fact that while $E_{J_1}$ has always the same analytical form, the energies of $J_0$ and $J_2$ SOC-states depend on relative strengths of the $J_\textup{H}$ and $\lambda$ as discussed earlier in Ref.~\cite{SI_kim2016prl}.

\begin{table}[h]
	\caption{Effective values of hopping $t_\textup{eff}$ estimated from DFT calculated values of bandwidth $W=2zt_\textup{eff}$ without SOC. $z$=12 for Ba$_2$YIrO$_6$, and $W$ has no effect of distortions in IrO$_6$ as $\Delta_{\textup{CF}}^{\textup{NC}}$= 0. For the 6H-iridates, $z$=2, and $\Delta_{\textup{CF}}^{\textup{NC}}$ also contributes.}
	\label{my-label}
	\begin{tabular}{|c|c|c|}
		\hline
		System              & Bandwidth $W$ (eV) & $t_\textup{eff}$ (eV) \\ \hline
		Ba$_2$YIrO$_6$      & 1.49               & 0.06                  \\ \hline
		Ba$_3$MgIr$_2$O$_9$ & 1.46               & 0.36                  \\ \hline
		Ba$_3$ZnIr$_2$O$_9$ & 1.22               & 0.31                  \\ \hline
		Ba$_3$CaIr$_2$O$_9$ & 1.38               & 0.34                  \\ \hline
		Ba$_3$SrIr$_2$O$_9$ & 1.23               & 0.31                  \\ \hline
	\end{tabular}
\end{table}

\begin{figure}[]
	\begin{center}
		\includegraphics[width=\columnwidth]{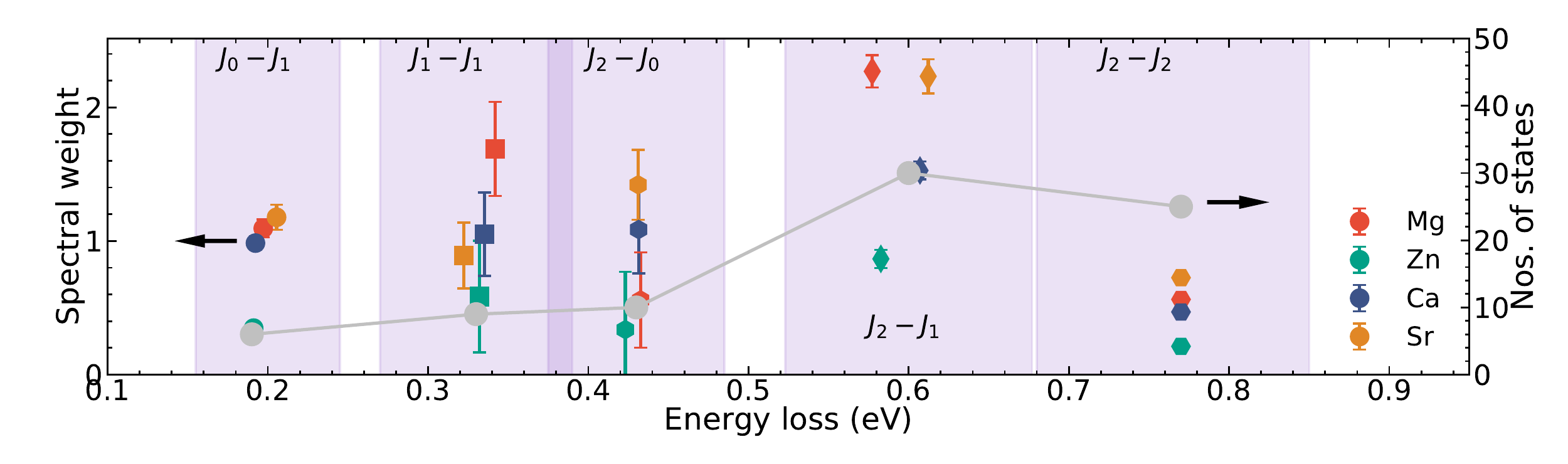}
		\caption{\textbf{Comparison of spectral weights of inelastic features.}
			Spectral weights of excitations to each set of two-site SOC states $J_0$-$J_1$, $J_1$-$J_1$, $J_2$-$J_0$ and $J_2$-$J_2$ are plotted, as obtained from the fits shown in Fig.~3 (a-d) of main text for the four 6H systems. Vertical bands are the FWHMs of the peaks obtained from RIXS on Ba$_3$ZnIr$_2$O$_9$. Using the right hand axis the number of two-site SOC states are plotted as gray dots and a line. For the excitation to the $J_2$-$J_2$ state (Fig.~3 (e) of main text), after subtraction of all other contributions, spectral weights are calculated by integrating between energy loss values from 0.7 to 0.8 eV. 
		}\label{SIfig5}
	\end{center}
\end{figure}
\section{Form of non-cubic crystal field matrices for the 6H-iridates}
\[
\mathrm{For~Mg~and~Zn,}~
H^{\Delta_\mathrm{CF}^\mathrm{NC}}=
\begin{bmatrix}
0 & 0 & 0\\
0 & 0 & 0\\
0 & 0 & \Delta\\
\end{bmatrix},~	
\mathrm{for~Ca,}~
\begin{bmatrix}
\Delta_1 & 0 & 0\\
0 & \Delta_2 & 0\\
0 & 0 & 0\\
\end{bmatrix},~
\mathrm{and~for~Sr,}~
\begin{bmatrix}
0 & 0 & 0\\
0 & \Delta_2 & 0\\
0 & 0 & \Delta_1\\
\end{bmatrix}.
\]

\section{Form of hopping matrices for the 6H-iridates}
\[
\mathrm{For~Mg~and~Zn,}~
H^t=
\begin{bmatrix}
-t_1 & 0 & 0\\
0 & -t_1 & 0\\
0 & 0 & -t_2\\
\end{bmatrix},~
\mathrm{for~Ca~and~Sr,}~
\begin{bmatrix}
t_1 & -t_2 & t_4\\
t_2 & -t_3 & t_4\\
-t_4 & t_4 & t_5\\
\end{bmatrix}
\]

\begin{table}[h]
	\caption{Ranges of physical parameters in the two-site model (equation (1) of main text and the matrices in Sections V and VI), using which the low-energy RIXS spectra could be fit for each system. The values given in italics are used for obtaining the energy-losses in Fig.~\ref{SIfig4}. We have kept $U_d$ as 2.0 eV for all calculations.}
	\label{rixsfit}
	\begin{tabular}{|c|c|c|c|c|}
		\hline
		System                               & $\Delta_{\textup{CF}}^{\textup{NC}}$ (eV) & $t_\textup{dim}$ (eV)                                                            & $J_\textup{H}$ (eV) & $\lambda$ (eV) \\ \hline
		\multirow{3}{*}{Ba$_3$MgIr$_2$O$_9$} & \multirow{3}{*}{$\Delta$={\it 0.025}}                    & \multirow{3}{*}{{\it $t_1$=0.16; $t_2$=0.041}}                                         & 0.35-0.46           & 0.26           \\ \cline{4-5}
		&                                           &                                                                                  & {\it 0.44-0.51}           & {\it 0.27}           \\ \cline{4-5}
		&                                           &                                                                                  & 0.48-0.55           & 0.28           \\ \hline
		Ba$_3$ZnIr$_2$O$_9$                  & $\Delta$={\it 0.02}                                      & {\it $t_1$=0.15; $t_2$=0.045}                                                          & {\it 0.42-0.49}           & {\it 0.26}          \\ \hline
		Ba$_3$CaIr$_2$O$_9$                  & $\Delta_2$={\it 0.044}, $\Delta_1$={\it  0.01}                               & {\it $t_1$=0.01; $t_2$=0.001; $t_3$=0.007; $t_4$=0.0; $t_5$=$t_3$}                     & {\it 0.45-0.5}            & {\it 0.26}           \\ \hline
		\multirow{2}{*}{Ba$_3$SrIr$_2$O$_9$} & \multirow{2}{*}{$\Delta_2$={\it 0.028}, $\Delta_1$={\it 0.01}}              & \multirow{2}{*}{{\it $t_1$=0.013; $t_2$=0.003; $t_3$=0.015; $t_4$=0.002; $t_5$=$t_3$}} & {\it 0.43-0.5}            & {\it 0.25}           \\ \cline{4-5}
		&                                           &                                                                                  & 0.54-0.58           & 0.26           \\ \hline
	\end{tabular}
\end{table}

\end{document}